\documentclass{aa}  

\usepackage{graphicx}
\usepackage{amsmath}
\usepackage{newtxtext,newtxmath}
\usepackage{multirow}

\usepackage{natbib,twoopt}
\usepackage[breaklinks=true]{hyperref}

\newcommand\ha            {H$\upalpha$}
\newcommand\hb            {H$\upbeta$}
\newcommand\nii	    	  {$\mathrm{\left[N\,\textsc{ii}\right] }$}
\newcommand\oi	    	  {$\mathrm{\left[O\,\textsc{i}\right] }$}
\newcommand\oiii      	  {$\mathrm{\left[O\,\textsc{iii}\right] }$}
\newcommand\sii	      	  {$\mathrm{\left[S\,\textsc{ii}\right]}$}

\newcommand\mha           {\mathrm{H}\alpha}

\newcommand\sbar          {\textsc{Sbar}}
\newcommand\wbar          {\textsc{Wbar}}
\newcommand\ubar          {\textsc{Ubar}}

\newcommand\fagn          {$f_{\mathrm{AGN}}$}
\newcommand\fagnsbar      {$f_{\mathrm{AGN, \sbar}}$}
\newcommand\fagnwbar      {$f_{\mathrm{AGN, \wbar}}$}
\newcommand\fagnubar      {$f_{\mathrm{AGN, \ubar}}$}

\newcommand\mmsun	     {\rm{M}_{\odot}}
\newcommand\km	         {\,\rm{km}}
\newcommand\yr	         {\,\rm{yr}}
\newcommand\s	         {\,\rm{s}}
\newcommand\Mpc	         {\,\rm{Mpc}}

\newcommand\mstar          {$M_{\ast}$}
\newcommand\mmstar         {M_{\ast}}
\newcommand\gmr            {$(g-r)_{0}$}

\newcommand\HST          {\emph{HST}}

\newcommand\euclid       {\emph{Euclid}}

\newcommand\astropy      {{\sc Astropy}}

\newcommand\eg          {\emph{e.g.},}
\newcommand\ie          {\emph{i.e.},}

\newcommand\refresp[1] {#1}

\defcitealias{garland2024}{G24}

\bibpunct{(}{)}{;}{a}{}{,}
\makeatletter
  \newcommandtwoopt{\citeads}[3][][]{\href{http://adsabs.harvard.edu/abs/#3}
    {\def\hyper@linkstart##1##2{}
     \let\hyper@linkend\@empty\citealp[#1][#2]{#3}}}
  \newcommandtwoopt{\citepads}[3][][]{\href{http://adsabs.harvard.edu/abs/#3}
    {\def\hyper@linkstart##1##2{}
     \let\hyper@linkend\@empty\citep[#1][#2]{#3}}}
  \newcommandtwoopt{\citetads}[3][][]{\href{http://adsabs.harvard.edu/abs/#3}
    {\def\hyper@linkstart##1##2{}
     \let\hyper@linkend\@empty\citet[#1][#2]{#3}}}
  \newcommandtwoopt{\citeyearads}[3][][]
    {\href{http://adsabs.harvard.edu/abs/#3}
    {\def\hyper@linkstart##1##2{}
     \let\hyper@linkend\@empty\citeyear[#1][#2]{#3}}}
\makeatother

\begin{document}

   \title{The complex relationships between AGN, bars and bulges}

   \author{I. L. Garland\inst{1,2}\fnmsep\thanks{\email{garland@mail.muni.cz}}
        \and H. Best\inst{1}
        \and L. F. Fortson\inst{3,4}
        \and T. G\'{e}ron\inst{5}
        \and C. J. Lintott\inst{6}
        \and D. O'Ryan\inst{7}
        \and B. D. Simmons\inst{2}
        \and R. J. Smethurst\inst{6}
        \and M. Viskotov\'{a}\inst{1}
        \and M. Walmsley\inst{5,8}
        \and N. Werner\inst{1}
        \and M. Zaja\v{c}ek\inst{1}
        }

   \institute{   Department of Theoretical Physics and Astrophysics, Faculty of Science, Masaryk University, Kotl\'{a}\v{r}sk\'{a} 2, Brno, 611 37, Czech Republic
            \and Department of Physics, Lancaster University, Lancaster, LA1 4YB, UK
            \and School of Physics and Astronomy, University of Minnesota, Minneapolis, Minnesota, 55455, USA
            \and Minnesota Institute for Astrophysics, University of Minnesota, Minneapolis, Minnesota, 55455, USA
            \and Dunlap Institute for Astronomy and Astrophysics, University of Toronto, 50 St. George Street, Toronto, ON M5S 3H4, Canada
            \and Oxford Astrophysics, Department of Physics, University of Oxford, Denys Wilkinson Building, Keble Road, Oxford, OX1 3RH, UK
            \and European Space Agency (ESA), European Space Astronomy Centre (ESAC), Camino Bajo del Castillo s/n, 28692, Villaneuva de la Ca\~{n}ada, Madrid, Spain
            \and Jodrell Bank Centre for Astrophysics, Department of Physics \& Astronomy, University of Manchester, Oxford Road, Manchester, M13 9PL, UK}

   \date{Received September 30, 20XX}

  \abstract
   {Via scaling relations, it is well-known that active galactic nuclei (AGN) and bulges are linked. This link was thought to be driven by mergers, but recent studies show that secular processes are the dominant mechanism of supermassive black hole growth. One such secular mechanism is gas inflow driven by large-scale bars. Since bulges can also grow via these bars, there is likely some common process between these three features.}
   {We investigate whether the observed correlation between AGN and bars is real or arises as a result of correlations between bars and bulges.} 
   {Using a catalogue of AGN identifications and galaxy morphologies in the DESI Legacy Survey at $z\leq0.1$, we control for mass and colour and investigate the AGN fraction variation with bulge prominence and bar strength.}
   {We first show that the variation in AGN fraction between strongly barred, weakly barred and unbarred galaxies does not qualitatively change if we additionally control for bulge prominence. Second, we find that in fixed bins of bulge prominence, the AGN fraction increases with increasing bar strength. In subsamples split by bar strength, the AGN fraction increases with bulge prominence, indicating that AGN presence correlates with both bar strength and bulge prominence simultaneously.}
   {}


   \maketitle
   \nolinenumbers

\section{Introduction}

The co-evolution of supermassive black holes (SMBHs) with their host galaxies is observed through a number of scaling relations \citep[see][for a review]{fabian2012, kormendy2013, heckman2014}.
SMBH masses have been found to correlate with both bulge properties, such as velocity dispersion and bulge stellar mass \refresp{\citep{ferrarese2000, haring2004, beifiori2012}}, and properties of the host galaxy as a whole, such as total stellar mass \citep{cisternas2011, marleau2013, simmons2017}.

SMBHs gain most of their mass during periods of rapid growth and accretion, during which they are observed as active galactic nuclei \citep[AGN;][]{shlosman1989}.
Therefore, by examining AGN, we can investigate the origins of this co-evolution between SMBHs and their host galaxies.

Whilst mergers between two or more galaxies are known to be one source of AGN triggering \citep[\eg][]{urrutia2008, glikman2015}, simulations have shown that most SMBH growth occurs via secular (\ie\ merger-free) pathways \citep{martin2018, mcalpine2020, smethurst2024}.
However, obtaining a pure and complete sample of galaxies with no major mergers in their recent history is highly challenging observationally.

\citet{martig2012} showed that galaxies with a bulge--to--total mass ratio of less than 0.1 (\ie\ little--to--no bulge) have had no mergers with a mass ratio greater than 1:4 since $z$$\sim$$2$.
Thus we could select bulgeless galaxies as a merger-free sample, however this is incomplete, since pseudobulges grow in the absence of mergers \citep{kormendy2004, kormendy2010}.
These look morphologically very similar to classical bulges, and without careful structural decomposition combined with dynamical analysis \citep[such as via the Kormendy Relation;][]{kormendy1977, hamabe1987}, distinguishing between secularly built pseudobulges and merger-built classical bulges is virtually impossible without significant cross-contamination between bulge types.
Additionally, there is substantial evidence for merger-free formation of classical bulges \citep{parry2009, bell2017, gargiulo2017, park2019, wang2019, guo2020, du2021}, for example through disk instabilities.
Thus, removing all galaxies with a bulge from a sample could mean removing a large number of secularly grown bulges.

\citet{simmons2017} use a sample of disk-dominated galaxies, without distinguishing bulge components between pseudobulge and classical bulge, and show that galaxy stellar mass correlates well with black hole mass in disk galaxies better than bulge stellar mass.
This sample was later confirmed \refresp{via careful structural decomposition and use of the Kormendy Relation} to have classical bulges in 53 per cent of the galaxies, and pseudobulges in 64 per cent, noting that some galaxies contain both a classical and a pseudobulge component \citep{fahey2025}.

The other crucial complication that arises when removing galaxies with a bulge component from a sample is that large-scale galactic bars can build up pseudobulges, providing a correlation between bar presence and bulge presence \citep{shlosman1989, kormendy2004, laurikainen2007, combes2009}.
Therefore, removing all galaxies with a bulge component would affect any observed relationship between bars and AGN.

A correlation between AGN presence and bar presence has been found in a number of works \refresp{\citep{knapen2000, laine2002, laurikainen2004a, coelho2011, oh2012, alonso2018, garland2023, kataria2024, lamarca2026}}.
However, due to the challenges in separating AGN emission from that of the host galaxy, along with the rarity of observationally merger-free disks (those with only a small bulge component), and the rarity of AGN, many of these studies find only a tenuous link, with high levels of uncertainty.
Other studies find no link at all \citep[\eg][]{cheung2015, goulding2017}.
Some studies \citep[\eg][]{galloway2015, silva-lima2022} find a higher AGN fraction in barred galaxies, but not higher levels of AGN activity, as measured by the \oiii\ luminosity.
\citet{garland2024} include all disk galaxies, regardless of their bulge size, and look at the AGN fraction with bar strength (divided into unbarred, strongly barred and weakly barred using Galaxy Zoo DESI machine-learning predicted volunteer votes) across the disk-dominated galaxy population.
In doing so, they show to a $>5\sigma$ confidence that strongly barred galaxies are more likely to host AGN than weakly barred galaxies, which are in turn more likely to host AGN than unbarred galaxies.

In this work, we investigate whether AGN presence correlates exclusively with bulge presence, or whether AGN presence is linked with both bars and bulges in some way.
We divide a sample of disk-dominated galaxies by bulge prominence (using Galaxy Zoo DESI), and investigate the AGN fraction in strongly barred, weakly barred and unbarred galaxies at each bulge prominence.
This allows us to test the AGN--bulge link at the same time as the AGN--bar link.
Since we do not have high quality photometric decomposition, along with measurements like surface brightness of each morphology component, we do not distinguish between classical and pseudobulges in this work.

This paper is structured as follows.
In Section \ref{sec:datacollation}, we discuss the sample selection.
Our results are presented in Section \ref{sec:results}, followed by discussion and conclusion in Sections \ref{sec:discussion} and \ref{sec:conclusions}.
Throughout this work, we use WMAP9 cosmology \citep{hinshaw2013}, where we assume a flat Universe, with $\mathrm{H}_{0}=69.3\km\s^{-1}\Mpc^{-1}$ and $\Omega_{m} = 0.287$, implemented via \astropy\ \citep{robitaille2013, price-whelan2018, price-whelan2022}.

\section{Sample Selection}\label{sec:datacollation}
In order to study the combined effect of galactic bulges and bars on AGN presence, we utilise the Galaxy Zoo: DESI catalogue \citep[GZD;][]{walmsley2023a}.
GZD consists of morphology classifications for 8.7 million galaxies in the DESI Legacy Surveys (DESI-LS), made with \emph{Zoobot}, a neural network trained on Galaxy Zoo volunteer votes \citep{walmsley2023b}.

In brief, DESI-LS consists of galaxies observed as part of DECaLS, BASS and MzLS\footnote{Dark Energy Camera Legacy Survey, Beijing-Arizona Sky Survey and Mayall z-band Legacy Survey respectively}.
Given the resulting size of DESI-LS, volunteer votes alone (as in previous Galaxy Zoo campaigns such as Galaxy Zoo 2 and Galaxy Zoo Hubble) are not efficient enough, and would take too long to collect for the entire catalogue.
Thus, volunteer votes on a subset of the data (401k galaxies) are used to train \emph{Zoobot}.
We refer the reader to the release paper for a detailed description of the initial catalogue.

To obtain the morphology and ionisation source classifications, we use the catalogue compiled in \citet[hereafter \citetalias{garland2024}]{garland2024}.
Again, we refer the reader to their paper for a detailed description, but summarise in brief here.

\citet{walmsley2023a} match GZD to the MPA-JHU SDSS DR7 catalogue \citep{abazajian2009} with a 3 arcsecond radius to obtain emission line fluxes, stellar masses and colours \citep{kauffmann2003, salim2007}.
\citetalias{garland2024} match to NYU-VAGC to obtain $k$-corrections \citep{blanton2005}, also within a 3 arcsecond radius.

In order to select a sample of not-edge-on, not-merging disks, \citetalias{garland2024} use the GZD model-predicted vote fractions.
\refresp{To select a sample of disks, \citetalias{garland2024} select galaxies with $f_{\mathrm{smooth-or-featured\_featured-or-disk}} \geq 0.27$, where $f_{\mathrm{smooth-or-featured\_featured-or-disk}}$ is the fraction of volunteers who voted for `featured or disk', as predicted by \emph{Zoobot}.
The cut-off value was recommended by \citet{walmsley2022}.
To select disks that are not edge-on, \citetalias{garland2024} again follow the procedure in \citet{walmsley2022} and select galaxies with $f_{\mathrm{disk-edge-on\_no}} \geq 0.68$, where $f_{\mathrm{disk-edge-on\_no}}$ is the model-predicted fraction of volunteers who voted for `not edge-on'.
In order to select merger-free galaxies, \citetalias{garland2024} define a new parameter of merger-prominence, $\zeta_{\mathrm{avg}}$, which combines the vote fractions for each category that volunteers can select -- `merging', `major disturbance', `minor disturbance', or `none'.
They identify $\zeta_{\mathrm{avg}} <0.3$ as being the ideal cut-off vote fraction to select merger-free galaxies, in order to maximise both completeness and purity.}
The first two conditions were described in \citet{walmsley2022}, and merger prominence in \citetalias{garland2024}.
\refresp{We want not-edge-on galaxies, since this makes bars (where present) easier to see, and whilst bars can sometimes be detected in edge-on galaxies, there is likely to be a bias present when asking volunteers to classify galaxies. Thus, to ensure we are removing this potential bias, we limit our sample to not-edge-on disks.}

Having compiled this initial sample, \citetalias{garland2024} separate the galaxies into unbarred, weakly barred, and strongly barred.
Using the methodology in \cite{geron2021}, a galaxy is designated as unbarred (\ubar) if $f_{\mathrm{strong-bar}}+f_{\mathrm{weak-bar}} < 0.5$, where $f_{\mathrm{x-bar}}$ is the model-predicted vote fraction for that bar strength.
Otherwise, it is considered barred.
This barred sample is then further split into strong and weak.
A galaxy is designated as weakly barred (\wbar) if it is not unbarred, and $f_{\mathrm{strong-bar}} < f_{\mathrm{weak-bar}}$.
A galaxy is designated as strongly barred (\sbar) if it is not unbarred, and $f_{\mathrm{strong-bar}} \geq f_{\mathrm{weak-bar}}$.

To ensure completeness and reduce selection effects, \citetalias{garland2024} volume-limit the sample, with redshift $z\leq0.1$, and $r$-band absolute magnitude $M_{r} \leq -19.2$, as shown in their Fig. 1.

Additionally, for this work we require an estimate of the bulge contribution to the galaxy morphology, which we refer to as bulge prominence.
\citet{masters2019} define a bulge prominence parameter, $B_{\mathrm{avg}}$, using SDSS morphology classifications from Galaxy Zoo 2 \citep[GZ2;][]{willett2013}.
However, GZ2 had only four different categories of bulge presence: none, just noticeable, obvious, and dominant.
GZD divides bulge presence into five categories: none, small, moderate, large and dominant.
Thus, we adapt $B_{\mathrm{avg}}$ to
\begin{align}
B = 0.25f_{\mathrm{small}} + 0.5f_{\mathrm{moderate}} + 0.75f_{\mathrm{large}} + 1.0f_{\mathrm{dominant}}
\end{align}
where $B$ is the bulge prominence parameter used in this work, and $f_{x}$ is the fraction of volunteers who voted for the bulge category $x$ as predicted by \emph{Zoobot}.

Note that there is no specific reason for these exact coefficients, as the aim is simply to condense the bulge vote fractions into one numeric parameter.
To confirm this, we tested several combinations of coefficients, and our results and analysis do not qualitatively change for any reasonable choice of coefficient weights on the different vote fractions.

As with any measurement, the GZD vote fractions do have errors associated with them.
When the vote fractions are varied within their errors (assumed to be Gaussian) using a bootstrapping method iterated 1000 times with replacements, our results do not qualitatively change.

\citetalias{garland2024} also publish classifications of the ionisation source.
The authors divide their sample via emission-line diagrams \citep{baldwin1981, veilleux1987, rosario2016} into AGN, star-forming, low-ionisation nuclear emission-line region (LINER), composite, undetermined, and uncertain.
The undetermined galaxies are those which have \ha\ flux with a signal-to-noise ratio of $\mathrm{S/N}_{\mha} < 3$, and thus have neither sufficient star-formation nor AGN activity to make an accurate determination.
Visual inspection shows that these undetermined galaxies (in this disk-dominated sample) are predominantly quiescent, red spirals.
Uncertain galaxies are those which are lacking sufficient signal-to-noise in other utilised emission lines (\hb, \oiii, \nii, \sii and \oi), such that they could theoretically fall into multiple other categories.
We remove from our sample uncertain galaxies (since their ionisation source remains unknown), composite galaxies (since the split between how much ionisation results from AGN compared to star-formation is unknown) and LINERs (since it remains debated whether these are low-luminosity AGN, or highly star-forming).
Again, we refer the reader to \citetalias{garland2024} for a full description of the ionisation classification procedure, notably their Fig. 2.

These cuts to the data result in our final volume-limited sample of 32\,683 disk-dominated, not edge-on, not merging galaxies that are either AGN, star-forming or undetermined.
There are 20\,417 unbarred galaxies, 9\,166 weakly barred, and 3\,100 strongly barred.
There are 3\,164 AGN hosts, 28\,807 star-forming galaxies (SFing), and 712 undetermined galaxies.
The median bulge prominence is 0.34, with a mean of 0.36 and a standard deviation of 0.07.

\section{Results}\label{sec:results}
\begin{figure*}
    \centering
	\includegraphics[width=\textwidth]{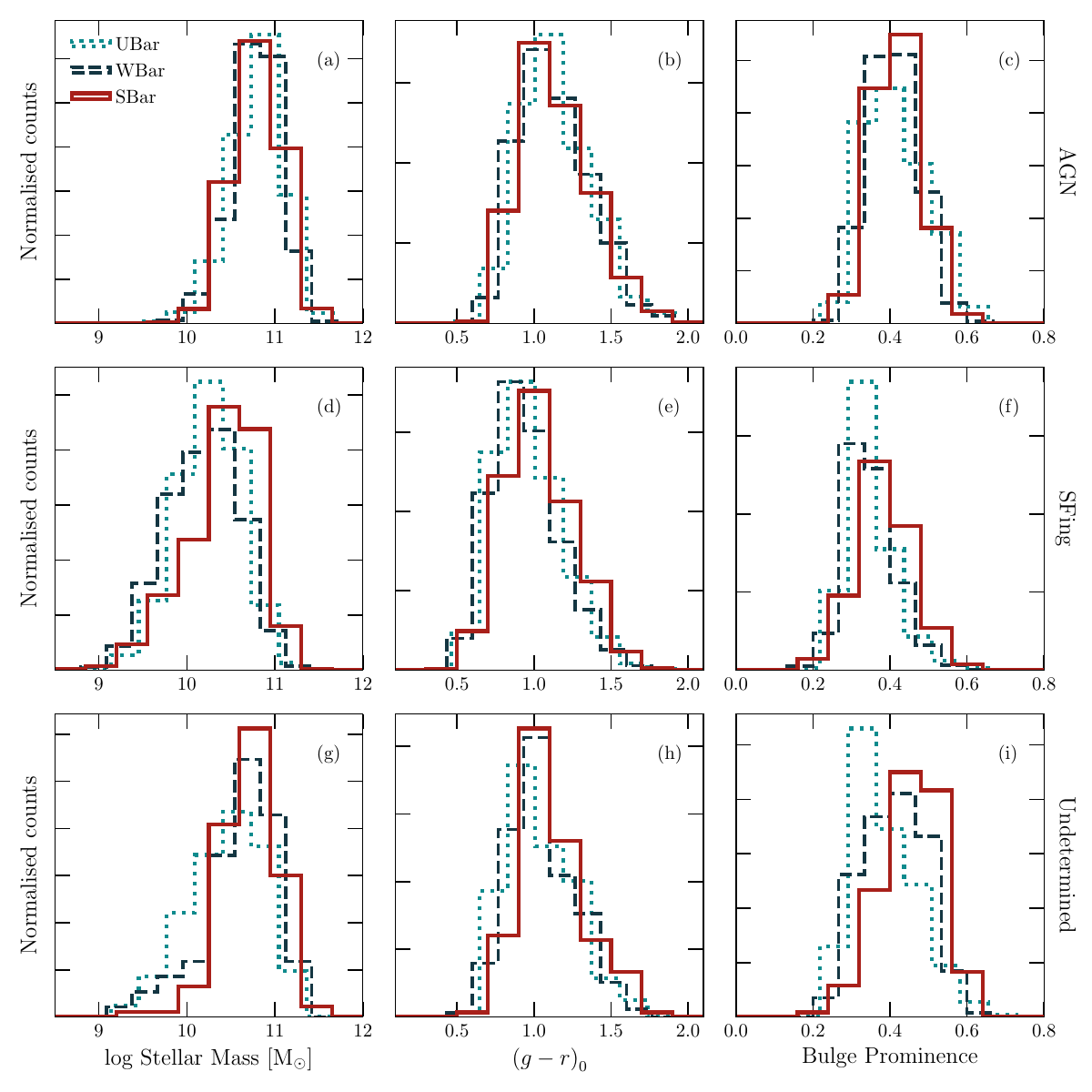}
    \caption{The distributions of stellar mass (left column), \gmr\ colour (middle column) and bulge prominence (right column) for AGN (top row), star-forming galaxies (middle row) and undetermined galaxies (bottom row).
    We show strongly barred galaxies in solid red lines, weakly barred in dashed navy blue, and unbarred in dotted teal.
    The AGN tend to have a higher bulge prominence, redder colour and higher stellar mass than their star-forming counterparts, although the ranges of these parameters do not vary significantly.
    The differences between the bar strengths are more apparent in star-forming galaxies than in AGN, with bulge prominence being particularly divided in undetermined galaxies.}
    \label{fig:oned_hists_used}
\end{figure*}

We first look at the spread of parameters thought to correlate with bar presence, and AGN presence: stellar mass, \refresp{k-corrected} \gmr\ colour (where the 0 indicates correction for Galactic absorption), and bulge prominence.
\refresp{The stellar masses and colours are taken from SDSS DR7, giving stellar mass errors on the order of $\Delta\log(\mmstar/\mmsun) \pm 0.1$.}
The distributions are shown in Fig. \ref{fig:oned_hists_used} for AGN, star-forming and undetermined sources, and in Fig. \ref{fig:oned_hists_additional} for LINERs, composite and uncertain sources, since we do not directly use the latter three in this work.

It is worth noting that the \refresp{smallest} bulge prominence \refresp{(\ie\ the low bulge-prominence end of the histogram)} in the AGN and the inactive (star-forming and undetermined) subsamples is very similar.
One might expect a higher minimum bulge prominence in the AGN subsample if the AGN gets mistaken for a bulge component during visual classifications.
Hence, we can be reassured that AGN-host galaxies are not mistakenly being labelled as bulge galaxies.

In order to account for the difference in \mstar, \gmr\ and $B$, we control for these three parameters via weighting our sample.
We divide our sample into 10 evenly-spaced bins \refresp{in each parameter, with cuts on the high and low ends of each parameter in order to remove outliers. These cuts are
\begin{itemize}
    \item $5.0 \leq \log(\mmstar/\mmsun) \leq 12.0$
    \item $-0.2 \leq (g-r)_{0} \leq 2.0 $
    \item $0 \leq B \leq 1.0$
\end{itemize}
Given that we are dividing each parameter into 10 bins, this gives us a total of 1000 bins.}

From here, \refresp{in each bin} we assign weights to each galaxy, such that the weighted distributions of these three parameters are the same between the \sbar, \wbar\ and \ubar\ subsamples.
\refresp{This means that our results with the weighted sample will not be affected by the mass, colour and bulge prominence.}
This extends the work of \citetalias{garland2024}, who only controlled for \mstar\ and \gmr.

\begin{figure}
    \centering
	\includegraphics[width=\columnwidth]{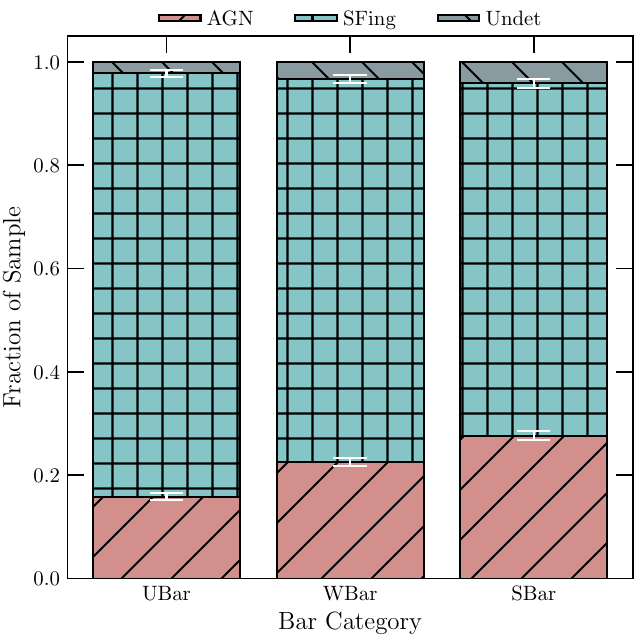}
    \caption{The fraction of galaxies in each bar strength that are AGN (red, positive diagonal), star-forming (SF; teal, square hatching) or undetermined (grey, negative diagonal).
    Error bars are shown in white.
    The AGN fraction increases as bar strength increases, although in each case the star-forming fraction is greater than the AGN fraction.}
    \label{fig:fagn_overall}
\end{figure}

We look at the overall AGN fraction (\fagn) in each of the bar subsamples.
These results are shown in Fig. \ref{fig:fagn_overall}, and Table \ref{tab:act_dist}.
After controlling for \mstar, \gmr\ and $B$, the AGN fraction in strongly barred galaxies is greater than that in weakly barred galaxies, which is greater than in unbarred galaxies, to $>3\sigma$ confidence.

\begin{table}
    \caption{The percentage of each ionisation category within each bar classification, as shown in Fig. \ref{fig:fagn_overall}.}
    \label{tab:act_dist}
    \begin{tabular}{llcccc}\hline
                                                  &       & \ubar\         & \wbar\         & \sbar\         &  \\ \hline
        \multirow{3}{*}{This work}                & AGN   & $15.9 \pm 0.7$ & $22.5 \pm 0.8$ & $27.6 \pm 0.8$ &  \\
                                                  & SFing & $81.9 \pm 0.7$ & $74.2 \pm 0.8$ & $68.2 \pm 0.9$ &  \\
                                                  & Undet & $ 2.2 \pm 0.3$ & $ 3.3 \pm 0.4$ & $ 4.2 \pm 0.4$ &  \\ \hline
        \multirow{3}{*}{\citetalias{garland2024}} & AGN   & $14.2 \pm 0.6$ & $23.3 \pm 0.8$ & $31.6 \pm 0.9$ &  \\
                                                  & SFing & $83.9 \pm 0.6$ & $73.6 \pm 0.8$ & $63.6 \pm 0.9$ &  \\
                                                  & Undet &  $1.9 \pm 0.2$ & $ 3.1 \pm 0.3$ & $ 4.7 \pm 0.4$ &  \\ \hline
    \end{tabular}
    \tablefoot{\ubar\ is unbarred galaxies, \wbar\ is weakly barred and \sbar\ is strongly barred.
    We have shown the results from \citetalias{garland2024} for comparison.
    AGN presence in strongly barred galaxies is around twice as prolific as in unbarred galaxies.}
\end{table}

Whilst the overall trends agree with those of \citetalias{garland2024}, the quantitative results differ slightly.
By controlling for bulge prominence, we still see that AGN fraction increases with bar strength, with a $>3\sigma$ difference between \fagnsbar, \fagnwbar\ and \fagnubar.
The quantitative values are in agreement with \citetalias{garland2024} to $3\sigma$ for \wbar\ and \ubar\ AGN fractions, but for \sbar, the AGN fraction is lower and the star-forming fraction is higher.
This implies that some of the observed difference in AGN fraction between strong and weakly barred galaxies is due to the bulge, but only a minority.
Even when controlling for bulge presence, the AGN fraction still increases with bar strength, as in \citetalias{garland2024}.

\begin{figure}
    \centering
	\includegraphics[width=\columnwidth]{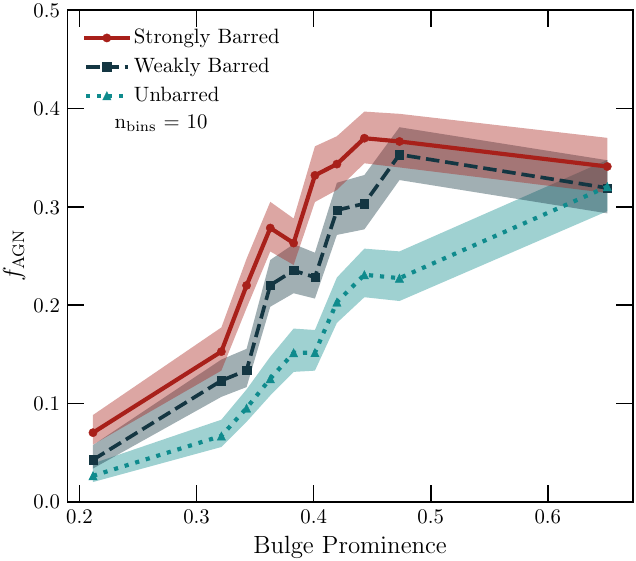}
    \caption{The effect of bulge prominence on AGN fraction (\fagn) for each of strongly barred (red solid line), weakly barred (navy dashed line) and unbarred (teal dotted line) disk galaxies.
    Overall, \fagn\ increases in each bar strength category with bulge prominence.
    At lower bulge prominences, \fagn\ increases in each bulge bin with bar strength, however the difference between \fagn\ in strongly and weakly barred galaxies disappears by $B\approx 0.48$, and the difference between all three bar categories disappears around $B\approx0.6$.
    The shaded regions show the $1\sigma$ uncertainties.}
    \label{fig:fagn_B_swubar}
\end{figure}

Given that much of the literature indicates a relationship between an AGN and the galactic bulge \citep[\eg\ through scaling relations, see][for a review]{kormendy2013}, we investigate how the AGN fraction changes with bulge prominence for each bar strength.
Using the sample described in Section \ref{sec:datacollation}, we control only for mass and colour as described above, using 10 bins for each.
We do not control for bulge prominence, since we want to investigate how AGN fraction changes with bulge prominence.
We divide our mass- and colour-controlled sample into 10 $B$ bins, such that each bin contains the same number of (weighted) galaxies.
Within each of these bins, we calculate the AGN fraction in strongly barred, weakly barred and unbarred galaxies.
The results are shown in Fig. \ref{fig:fagn_B_swubar}.

Interestingly, we do see an overall increase in each bar strength of AGN fraction with bulge prominence, \ie\ within a specific bar category, the AGN fraction increases overall with bulge prominence.
We also see that within each bulge prominence bin, the AGN fraction increases with bar strength.
However at higher bulge prominences, the picture becomes less clear, with the difference between strong and weak bars fading at around $B\approx0.45$, and the differences between all bar categories fading around $B\approx0.6$.

\begin{figure}
    \centering
	\includegraphics[width=\columnwidth]{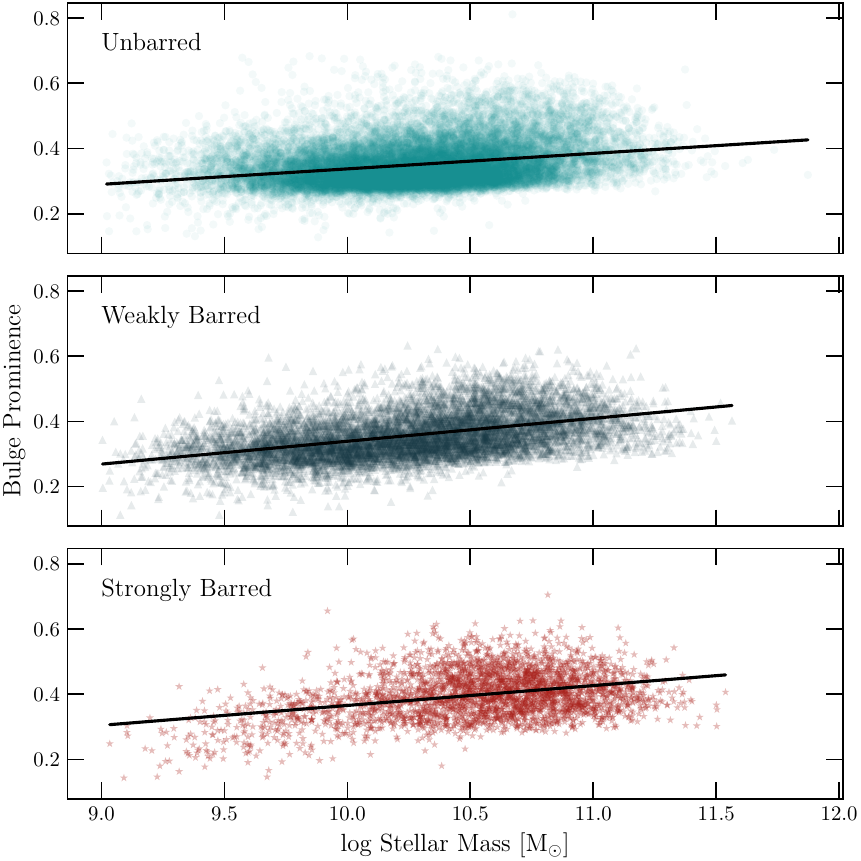}
    \caption{The relationship between stellar mass and bulge prominence for our sample.
    Lines of best fit are shown in black. We split the sample by unbarred (teal), weakly barred (navy blue) and strongly barred (red) galaxies.
    There is a slight increase in the bulge prominence with stellar mass.}
    \label{fig:B_mass}
\end{figure}

We want to ensure that we are not just seeing a trend with stellar mass in Fig. \ref{fig:fagn_B_swubar}, since bulge prominence can vary with stellar mass \citep[\eg][]{huertascompany2025}.
In order to negate the effect of stellar mass, we follow a similar methodology to that used in \citet{masters2012}, although instead of their bar fraction, we use AGN fraction, and instead of their gas fraction, we use bulge prominence.
We show the relationship between bulge prominence and stellar mass for our (uncontrolled) sample in Fig. \ref{fig:B_mass}.

Trends are seen for each bar strength.
We use linear regression to show that the line of best fit for each bar category is

\begin{align}
    \langle B_{\mathrm{\ubar}}\rangle &= 0.070\log(\mmstar/\mmsun) - 0.387 \\
    \langle B_{\mathrm{\wbar}}\rangle &= 0.077\log(\mmstar/\mmsun) - 0.445 \\
    \langle B_{\mathrm{\sbar}}\rangle &= 0.042\log(\mmstar/\mmsun) - 0.058
\end{align}

\begin{figure}
    \centering
	\includegraphics[width=\columnwidth]{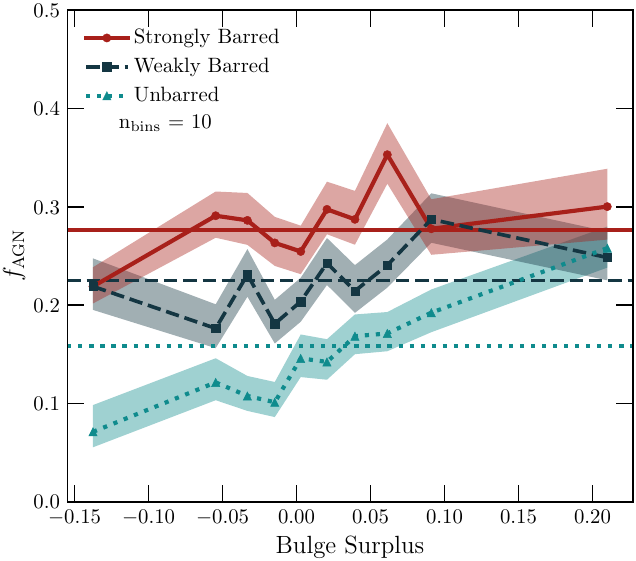}
    \caption{Variation of \fagn\ with the bulge surplus, as calculated in Equation \ref{eq:bulge_surplus}. The relationship for strongly barred galaxies is shown in solid red, for weakly barred in dashed navy blue, and for unbarred in dotted teal.
    The horizontal lines show the mean \fagn\ for each bar category.
    Shaded regions show the $1\sigma$ uncertainties.
    Galaxies at a given stellar mass are more likely to be hosting an AGN if they also have a greater bulge surplus, and this relationship is steeper for unbarred galaxies than for strongly or weakly barred.}
    \label{fig:fagn_b_surplus}
\end{figure}

From here, we can define a measure of bulge surplus, $B_{\mathrm{surp}}$ \ie\ how much higher a bulge prominence does a galaxy have for a given stellar mass, 

\begin{align}\label{eq:bulge_surplus}
    B_{\mathrm{surp}} = B - \langle B_{\mathrm{\textsc{Xbar}}}\rangle
\end{align}

where \textsc{Xbar} represents the relevant bar category.
We then plot the AGN fraction in each bar strength with the bar surplus, using our mass- and colour-controlled sample.
The results are shown in Fig. \ref{fig:fagn_b_surplus}.

The horizontal lines show the median \fagn\ for each bar strength.
There is a slight increase in the AGN fraction as the bulge surplus increases.
In other words, if the bulge is more prominent than expected for its host galaxy's stellar mass, then there is more likely to be an AGN.
This trend is stronger for unbarred galaxies than strongly barred.

\begin{figure*}
    \centering
	\includegraphics[width=\textwidth]{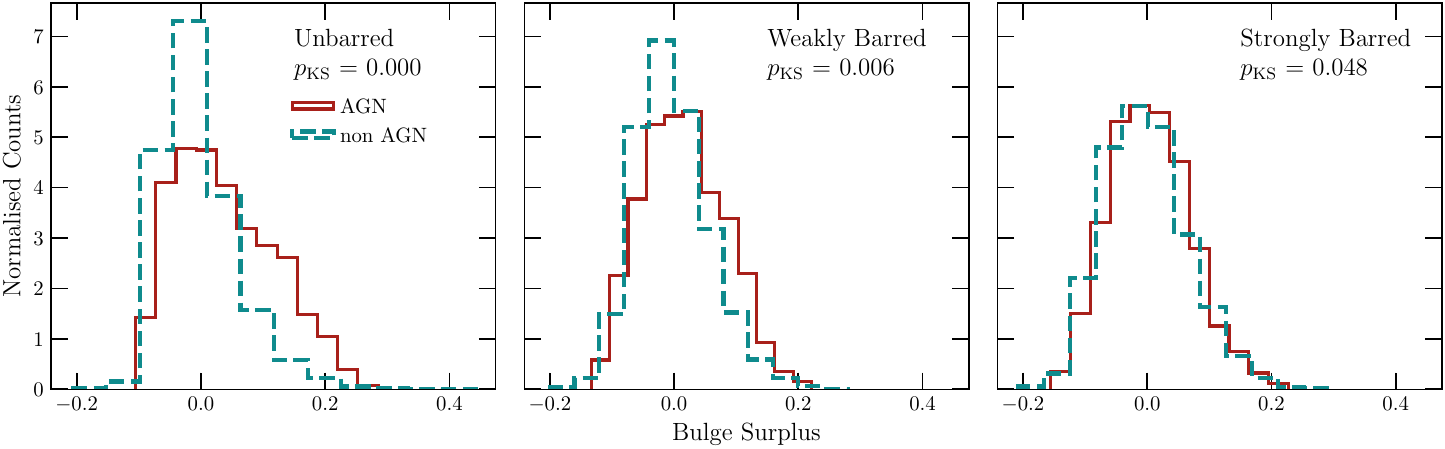}
    \caption{The normalised distributions of bulge surplus for unbarred (left), weakly barred (centre) and strongly barred (right) galaxies, split between AGN (red) and non-AGN host galaxies (teal).
    We compare the AGN and non-AGN distributions in each bar category using KS tests, and the p-values are shown on the relevant plots.}
    \label{fig:b_surplus_hists}
\end{figure*}

We can also quantify the difference in bulge surplus for AGN versus inactive galaxies via a KS test \citep{kolmogorov1933}. The histograms of the bulge surplus distribution are shown in Fig. \ref{fig:b_surplus_hists}, with $p$-values for the KS tests written on the plots.

For weakly barred and strongly barred galaxies, the bulge surplus distribution for AGN and inactive galaxies are consistent with being drawn from the same parent sample ($2.74\upsigma$ and $1.97\upsigma$ respectively).
The unbarred bulge surplus distributions for AGN and inactive galaxies are inconsistent with being drawn from the same parent sample ($>5\upsigma$).
This indicates that in unbarred galaxies, the excess bulge component is likely linked to AGN presence, but such a bulge component makes less difference in barred galaxies.
\refresp{When we limit the redshift to $z$$\sim$0.05, the difference in bulge surplus between the unbarred AGN and unbarred non AGN lessens, giving $p=0.08$, however this is likely due to the significant decrease in unbarred sample size by a factor of 5. The p-values for strongly and weakly barred do not significantly change. Further work with high resolution images will determine if this is a real trend or a selection effect.}

\section{Discussion}\label{sec:discussion}
The positive correlations between AGN fraction (\fagn) and bulge prominence ($B$) and between \fagn\ and bar strength in Fig. \ref{fig:fagn_B_swubar} indicate that there is a highly complex interplay between these three features.
There is not only one correlation that mimics the other, and AGN presence correlates with both bar strength and bulge prominence even when controlling for the other.

This indicates that AGN can be triggered and/or fuelled both in galaxies with and without a bulge, with there being a higher AGN fraction in galaxies with a bulge.
However, at every bulge prominence, there is a higher AGN fraction in strongly barred galaxies.
Similarly, AGN can be triggered and/or fuelled in galaxies with strong bars, weak bars or no bars, with there being a higher AGN fraction in strongly barred galaxies.
However, at every bar strength, the AGN fraction increases with bulge prominence.

Scaling relations have long demonstrated a link between AGN and bulge properties (\ie\ the \citet{haring2004} relationship between black hole mass and bulge stellar mass), however these only discuss the connection between AGN that are already switched on, not the presence of the AGN itself.
Thus, we know that black hole mass is related to bulge mass, but this does not necessarily mean that bulges are responsible for the switching on of an AGN.
Our work however, shows that AGN fraction increases with bulge prominence -- larger bulges are more likely to host an AGN, indicating that the bulge size (relative to the host galaxy) is linked to AGN switch-on.
Note, however, that many of the bulges in our galaxy sample are likely secularly grown, due to the disk dominated nature of the galaxies.

The relationship between bars and AGN is less well understood.
Recent works, \refresp{including simulations such as \citet{kataria2024, frosst2025} and observations such as \citet{silva-lima2022, garland2024}} indicate that AGN are more likely to lie in galaxies with a bar, in agreement with our results.
However works such as \refresp{\citet[via X-ray selected AGN]{goulding2017} and \citet[via analysis of AGN and non-AGN barred galaxies separately]{zee2023}} show no such correlation.
\citet{marels2025} show that AGN in barred galaxies are more powerful than in unbarred galaxies, although they do not discuss the AGN presence, similar to the scaling relations described above.

Bars can grow bulges over time via funnelling gas into the centre of the galaxy \citep[\eg][]{combes2009}.
Our results indicate that if a bar is sufficient to grow a bulge component, it is also sufficient to trigger an AGN.

The tapering off of an AGN fraction \refresp{(\ie\ where the graph begins to level out at around 40 per cent) }in both \refresp{of the} barred subsamples is likely due to the AGN duty cycle.
AGN seemingly flicker on and off over their lifetime \citep{schawinksi2015}, so this indicates that the ``on'' accounts for around 40 per cent of the AGN total lifetime.
The unbarred AGN fraction may also peak around this point at higher bulge prominence, but we do not have sufficient data to inform this.
The strongly barred galaxies reaching this plateau at a lower $B$ than weakly barred is indicative that strongly barred galaxies are fuelling AGN more effectively than weakly barred, which need to build up a higher bulge prominence before triggering AGN switch-on.
This could mean that weak bars take longer to trigger an AGN.

We propose the following duty cycle.
Assume that there is a galaxy with a disk, no bulge or bar component, and an inactive SMBH at its centre.
Such a disk, after some time, forms a bar either through buckling instability, or through a tidal interaction \citep[\eg][]{hohl1971, noguchi1987, sellwood1993, skibba2012}.
This bar would funnel gas to the centre of the galaxy, triggering an AGN and forming a bulge \citep[\eg][]{kormendy2004, athanassoula2005, laurikainen2007, combes2009}.
The stronger the bar is, the more likely it is to trigger the switch-on of an AGN \citepalias{garland2024}.
Simultaneously, the bar can also build up a bulge component, thus meaning that bars that are funnelling enough gas to develop a bulge are also likely to trigger an AGN.
If there is a sufficient gas supply, then the bar will allow the bulge to increase in size.
At some point, the gas supply runs out, and the AGN switches off, leading to a maximum \fagn\ of around 40 per cent.
If the galaxy never develops a bar, its time spent as an AGN that we can detect depends on its bulge strength, but the influence of the bar dominates over this effect, if it is present.

\refresp{Further work will be needed to understand whether such a duty cycle effect is really in operation, including how the AGN fraction varies in barred galaxies over a greater range of redshifts, and whether this effect is stronger in a particular stellar mass regime.}

Longslit spectroscopic data \refresp{or integral field unit (IFU) spectra }could allow us to measure the ages of the stellar populations in the bar and bulge separately in AGN hosts, rather than relying on the prominence of the bulge as a proxy for the age of the bar.
This would help us to confirm or rule out the proposed duty cycle\refresp{, however should be combined with high resolution imagery such as that from \euclid\ in order to account for other structures, such as nuclear rings and nuclear bars}.

The other key reason to consider bar age is that bars are often much longer-lived structures than AGN -- $\sim$$10^{9-10}\yr$ for bars \citep{kraljic2012, sellwood2014} compared to $\sim$$10^{5}\yr$ for AGN phases \citep{schawinksi2015}.
Where we see a barred galaxy without an AGN, it could be that the bar did trigger an AGN that has since switched off.
This would be true of other processes as well, and is not just a caveat for studies investigating bar-driven growth.

It is highly important to consider our selection effects when drawing conclusions.
The galaxies used in this sample are part of the DESI Legacy Survey, which requires that the point-spread function of an image in the $z$-band is a maximum of 1.5 arcseconds.
At the redshifts of this work ($z \leq 0.1$), this is equivalent to 2.766 kpc.
Any bulges or bars smaller than this may not be resolved, and thus will remain undetected.
Higher-resolution photometry (\eg\ from \HST\ or \euclid) is required to pick out these smaller components and make more accurate morphology classifications.
However, despite the limitation on photometry, \citet{fahey2025} showed that samples can be selected from ground-based surveys such as SDSS that are later confirmed to be disk-dominated with \HST\ photometry.

\section{Conclusions}\label{sec:conclusions}
We have used the Galaxy Zoo: DESI catalogue first presented in \citet{walmsley2023a} and the classifications first presented in \citet{garland2024} to investigate the dual effect of bulge prominence and bar strength on AGN presence.
Our key results can be summarised as follows:
\begin{itemize}
    \item After controlling for bulge prominence, as well as stellar mass and \gmr, we find that the AGN fractions in subsamples split by bar strength are in excellent agreement with \citet{garland2024}, where they only controlled for stellar mass and \gmr.
    That is, that strongly barred galaxies have a higher AGN fraction than weakly barred, which have a higher AGN fraction than unbarred.
    \item When we split our controlled sample into bins of bulge prominence, we find the same trends -- more strongly barred subsamples have a higher AGN fraction.
    \item We propose a duty cycle linking the activity of the AGN, bar and bulge, wherein the bar triggers the AGN to switch on, whilst simultaneously building up a bulge.
\end{itemize}

Further work is required to investigate these AGN that are fuelled in the absence of bulge components or bar components, as well as investigation of the inactive galaxies where there is a bar and/or bulge present.
IFU data, or longslit spectroscopic data at multiple angles would allow us to measure the ages of stellar populations in the bar and bulge, as well as measure gas content with respect to the morphological components.
Large-scale surveys at high resolution, such as those being conducted by \euclid, would allow for parametric decomposition of galaxies to a high precision, allowing us to identify bar strength and bulge prominence to a higher confidence.

\begin{acknowledgements}
ILG, HB, MV and MZ have received the support from the Czech Science Foundation Junior Star grant no. GM24-10599M (``Stars in galactic nuclei: interrelation with massive black holes'').
LFF acknowledges partial support from NASA Award \#80NSSC20M0057.
BDS acknowledges support through a UK Research and Innovation Future Leaders Fellowship [grant number MR/T044136/1] and its renewal [grant number MR/Z000076/1].
TG is a Canadian Rubin Fellow at the Dunlap Institute.
CJL acknowledges support from the Sloan Foundation.
RJS gratefully acknowledges support through the Royal Astronomical Society Research Fellowship.
MW is a Dunlap Fellow. The Dunlap Institute is funded through an endowment established by the David Dunlap family and the University of Toronto.

The data in this paper are the result of the efforts of the Galaxy Zoo volunteers, without whom none of this work would be possible.
Their efforts are individually acknowledged at \url{http://authors.galaxyzoo.org}.

This research has used TOPCAT \citep{taylor2005}, an interactive graphical tool for analysis and manipulation of tabular data.

This research has made extensive use of the following Python packages: \textsc{Astropy}, a community-developed core Python package for Astronomy \citep{robitaille2013, price-whelan2018, price-whelan2022}; \textsc{Matplotlib}, a 2D graphics package for Python \citep{hunter2007}; \textsc{Numpy} \citep{harris2020}, a package for scientific computing; \textsc{Scipy} \citep{virtanen2020}, a package for fundamental algorithms in scientific computing.

Funding for the SDSS and SDSS-II has been provided by the Alfred P. Sloan Foundation, the Participating Institutions, the National Science Foundation, the U.S. Department of Energy, the National Aeronautics and Space Administration, the Japanese Monbukagakusho, the Max Planck Society, and the Higher Education Funding Council for England.
The SDSS Web Site is \url{http://www.sdss.org/}.

The SDSS is managed by the Astrophysical Research Consortium for the Participating Institutions.
The Participating Institutions are the American Museum of Natural History, Astrophysical Institute Potsdam, University of Basel, University of Cambridge, Case Western Reserve University, University of Chicago, Drexel University, Fermilab, the Institute for Advanced Study, the Japan Participation Group, Johns Hopkins University, the Joint Institute for Nuclear Astrophysics, the Kavli Institute for Particle Astrophysics and Cosmology, the Korean Scientist Group, the Chinese Academy of Sciences (LAMOST), Los Alamos National Laboratory, the Max-Planck-Institute for Astronomy (MPIA), the Max-Planck-Institute for Astrophysics (MPA), New Mexico State University, Ohio State University, University of Pittsburgh, University of Portsmouth, Princeton University, the United States Naval Observatory, and the University of Washington.

The Legacy Surveys consist of three individual and complementary projects: the Dark Energy Camera Legacy Survey (DECaLS; Proposal ID \#2014B-0404; PIs: David Schlegel and Arjun Dey), the Beijing-Arizona Sky Survey (BASS; NOAO Prop. ID \#2015A-0801; PIs: Zhou Xu and Xiaohui Fan), and the Mayall z-band Legacy Survey (MzLS; Prop. ID \#2016A-0453; PI: Arjun Dey).
DECaLS, BASS and MzLS together include data obtained, respectively, at the Blanco telescope, Cerro Tololo Inter-American Observatory, NSF’s NOIRLab; the Bok telescope, Steward Observatory, University of Arizona; and the Mayall telescope, Kitt Peak National Observatory, NOIRLab.
Pipeline processing and analyses of the data were supported by NOIRLab and the Lawrence Berkeley National Laboratory (LBNL).
The Legacy Surveys project is honored to be permitted to conduct astronomical research on Iolkam Du’ag (Kitt Peak), a mountain with particular significance to the Tohono O’odham Nation.

NOIRLab is operated by the Association of Universities for Research in Astronomy (AURA) under a cooperative agreement with the National Science Foundation.
LBNL is managed by the Regents of the University of California under contract to the U.S. Department of Energy.

This project used data obtained with the Dark Energy Camera (DECam), which was constructed by the Dark Energy Survey (DES) collaboration.
Funding for the DES Projects has been provided by the U.S. Department of Energy, the U.S. National Science Foundation, the Ministry of Science and Education of Spain, the Science and Technology Facilities Council of the United Kingdom, the Higher Education Funding Council for England, the National Center for Supercomputing Applications at the University of Illinois at Urbana-Champaign, the Kavli Institute of Cosmological Physics at the University of Chicago, Center for Cosmology and Astro-Particle Physics at the Ohio State University, the Mitchell Institute for Fundamental Physics and Astronomy at Texas A\&M University, Financiadora de Estudos e Projetos, Fundacao Carlos Chagas Filho de Amparo, Financiadora de Estudos e Projetos, Fundacao Carlos Chagas Filho de Amparo a Pesquisa do Estado do Rio de Janeiro, Conselho Nacional de Desenvolvimento Cientifico e Tecnologico and the Ministerio da Ciencia, Tecnologia e Inovacao, the Deutsche Forschungsgemeinschaft and the Collaborating Institutions in the Dark Energy Survey.
The Collaborating Institutions are Argonne National Laboratory, the University of California at Santa Cruz, the University of Cambridge, Centro de Investigaciones Energeticas, Medioambientales y Tecnologicas-Madrid, the University of Chicago, University College London, the DES-Brazil Consortium, the University of Edinburgh, the Eidgenossische Technische Hochschule (ETH) Zurich, Fermi National Accelerator Laboratory, the University of Illinois at Urbana-Champaign, the Institut de Ciencies de l’Espai (IEEC/CSIC), the Institut de Fisica d’Altes Energies, Lawrence Berkeley National Laboratory, the Ludwig Maximilians Universitat Munchen and the associated Excellence Cluster Universe, the University of Michigan, NSF’s NOIRLab, the University of Nottingham, the Ohio State University, the University of Pennsylvania, the University of Portsmouth, SLAC National Accelerator Laboratory, Stanford University, the University of Sussex, and Texas A\&M University.

BASS is a key project of the Telescope Access Program (TAP), which has been funded by the National Astronomical Observatories of China, the Chinese Academy of Sciences (the Strategic Priority Research Program “The Emergence of Cosmological Structures” Grant \# XDB09000000), and the Special Fund for Astronomy from the Ministry of Finance.
The BASS is also supported by the External Cooperation Program of Chinese Academy of Sciences (Grant \# 114A11KYSB20160057), and Chinese National Natural Science Foundation (Grant \# 12120101003, \# 11433005).

The Legacy Survey team makes use of data products from the Near-Earth Object Wide-field Infrared Survey Explorer (NEOWISE), which is a project of the Jet Propulsion Laboratory/California Institute of Technology.
NEOWISE is funded by the National Aeronautics and Space Administration.

The Legacy Surveys imaging of the DESI footprint is supported by the Director, Office of Science, Office of High Energy Physics of the U.S. Department of Energy under Contract No. DE-AC02-05CH1123, by the National Energy Research Scientific Computing Center, a DOE Office of Science User Facility under the same contract; and by the U.S. National Science Foundation, Division of Astronomical Sciences under Contract No. AST-0950945 to NOAO.
\end{acknowledgements}

\bibliographystyle{aa}
\bibliography{bibliography}

\begin{appendix}

\onecolumn
\section{Supplementary stellar mass, colour and bulge distributions}\label{sec:additional_mass_gmr_hists}
\begin{figure*}[h!]
    \centering
     \resizebox{12cm}{12cm}
    {\includegraphics{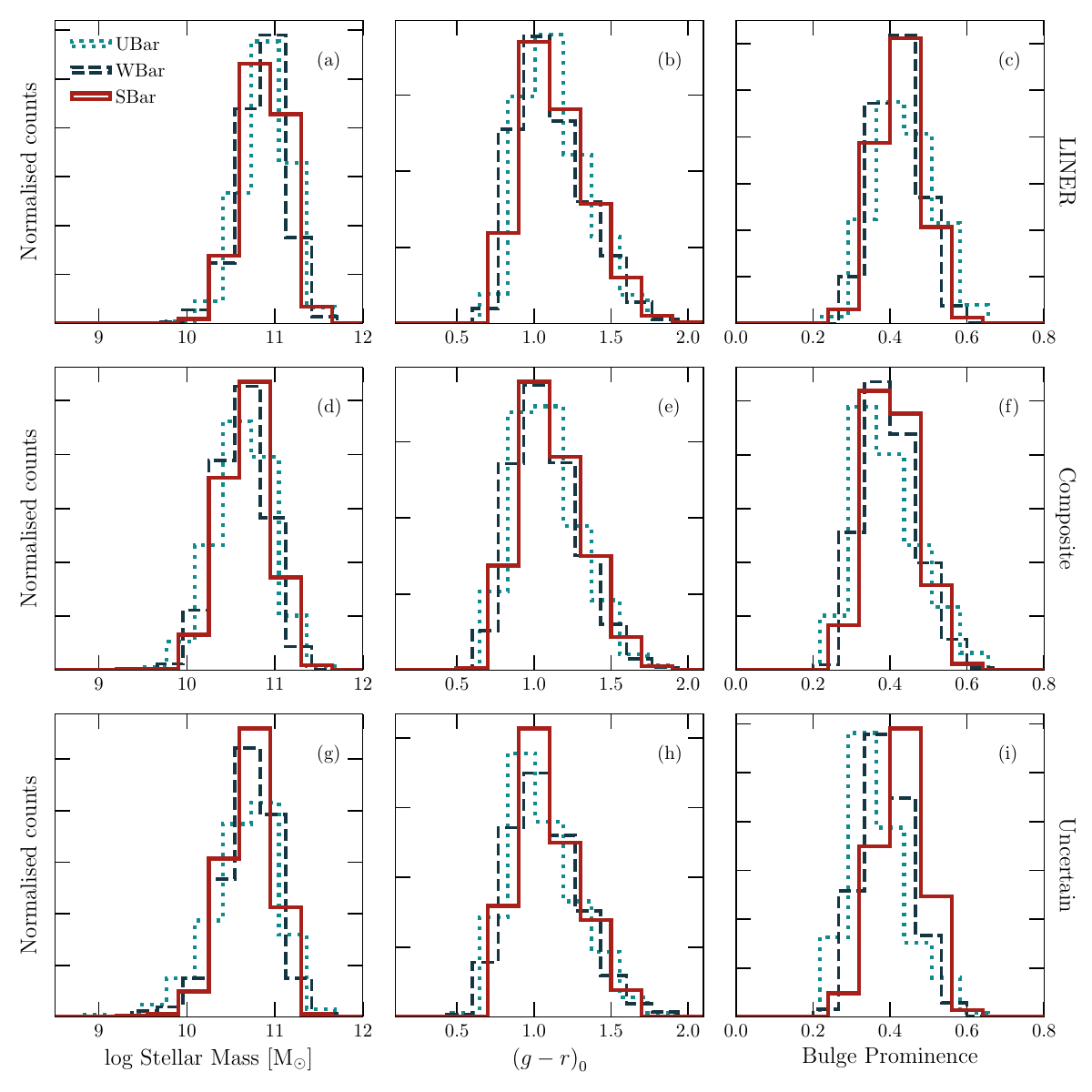}}
     \caption{The distributions of stellar mass (left column), \gmr\ colour (middle column) and bulge prominence (right column) for LINER (top row), composite galaxies (middle row) and uncertain galaxies (bottom row).
     We show strongly barred galaxies in solid red lines, weakly barred in dashed navy blue, and unbarred in dotted teal.}
      \label{fig:oned_hists_additional}
\end{figure*}

Fig. \ref{fig:oned_hists_additional} shows the stellar mass (\mstar), \gmr\ colour and bulge prominence ($B$) distributions for the LINERs, composite galaxies and uncertain in our sample, in a matter identical to Fig. \ref{fig:oned_hists_used}.

\end{appendix}
\end{document}